# Characterization of the new free-air primary standard for medium-energy X-rays at CMI

**Vladimir Sochor and Jaroslav Solc***

*Czech Metrology Institute, Okruzni 31, 638 00 Brno, Czech Republic*
*\* corresponding author; e-mail*: `jsolc@cmi.cz`

ABSTRACT: In 2011 a decision was made by Czech Metrology Institute to build a free-air ionization chamber (FAC) intended to be used as a primary standard of air kerma rate for medium-energy X-rays (photon energy from 40 to 300 keV, including mammography X-ray qualities) in order to replace currently used secondary ionization chamber and to decrease the uncertainty of air kerma reference value. In period of 2013-2017, the FAC has been designed, manufactured and put into operation. Correction factors were measured or calculated by a Monte Carlo method. FAC performance was preliminary tested using a calibrated secondary chamber. Physical characteristics of the FAC are described and a summary of the correction factors with the uncertainty budget is presented.

**Contents**

**1. Introduction**

Czech Metrology Institute (CMI) provides the traceability for measurement instruments in different qualities of X-rays (including mammography) by means of Czech national standard of exposure and air kerma No. ECM 440-5/11-049. This national standard has been built as a secondary standard, i.e. it is based on a set of secondary cavity ionizing chambers covering all basic photon energy regions (low-energy X-rays up to 33 keV, medium energy X-rays 33-250 keV and gamma radiation of Cs-137 and Co-60). The chambers are traceable to primary standards abroad. In 2011 CMI made a decision to commence building primary standards covering the energy range of photons available in CMI. During 2011 and 2012 the free-air ionization chamber (FAC) for low-energy X-rays (up to approximately 50 keV) was built [1] which in 2015 successfully participated in key comparisons [2],[3]. Based on experiences with the low-energy FAC, this development continued in 2013 – 2017 with a larger FAC designed for medium-energy X-rays up to 300 keV that is described in this paper. In 2015 – 2017 CMI participated in the European Metrology Programme for Innovation and Research (EMPIR) joint research project 14RPT04 "Absorbed dose in water and air" focused to primary standards of air kerma and absorbed dose to water. Experience gained during the project from other partners was a new valuable source of information supporting the FAC development and manufacture.

**2. Materials and methods**

**2.1 Air kerma determination**

Value of air kerma, $K_{air}$, at the entrance opening can be determined from measured value of ionizing current according to equations (1) to (3).

$$K_{air} = \frac{Q}{A.l.\rho} \frac{W_{air}}{e} \frac{\prod k_i}{(1-g)} \tag{1}$$

or

$$\dot{K}_{air} = N_{K,air}.I.\prod k_i \tag{2}$$

where

$$N_{K,air} = \frac{K_{air}}{Q} = \frac{W_{air}}{e} \frac{1}{A.l.\rho.(1-g)} \tag{3}$$

In equations (1) to (3), $Q$ is measured electric charge, $A$ is area of entrance aperture, $l$ is effective length of collecting electrode, $\rho$ is air density, $W_{air}/e$ is energy needed to create an ion pair in dry air [4] ($W_{air}/e$ = 33,97 J/C), $g$ is a fraction of energy lost due to bremsstrahlung [5] (for X-ray photons in a region from 8 keV to 300 keV $g$ is negligible ($g$ = 0)), $k_i$ are the necessary correction factors: $k_{ii}$ – correction for a difference between the number of collected ions and collected charge (initial-ionization correction) [4], $k_W$ – correction for the change of the $W_{air}$ value for lower photon energy [4], $k_h$ – correction for air humidity [6], $k_{tp}$ – correction for air pressure and temperature, $k_{att}$ – correction for air attenuation between the entrance opening and collecting electrode, $k_{sc}$ – correction for scattered photons, $k_{fl}$ – correction for fluorescence photons created in argon in air, $k_{loss}$ – correction for loss of charge, which doesn't contribute to the signal from collecting electrode, $k_d$ – correction for an electric field deterioration, $k_{pol}$ – correction for polarity, $k_{sat}$ – correction for saturation or ion recombination and diffusion, $k_{ap}$ – correction for influence of entrance opening, i.e. influence of photons passing through the entrance diaphragm $k_{dtr}$ and influence of photon scattered on the edge of entrance opening $k_{dsc}$ ($k_{ap} = k_{dtr} \cdot k_{dsc}$), $k_{tr}$ – correction for photons passing through the chamber's front wall. Meanings of individual correction factors are described in [7] in more detail.

## 2.2 FAC design

Design of the FAC of CMI is based on similar designs in different primary laboratories. Basic dimensions of the chamber ensure that the chamber can be used for photons of energy up to approximately 300 keV. FAC is built in usual way as a plan-parallel ionization chamber (Figures 1 and 2). Beam of X-rays is entering the shielded metal box through an opening of an area $A$, then it is traveling between two plan-parallel electrodes and leaves the box through an opening on the opposite side without touching the structure of the FAC. Between these electrodes an electric field exists, its homogeneity is maintained by a set of guarding rings connected to equidistantly divided polarizing potential.

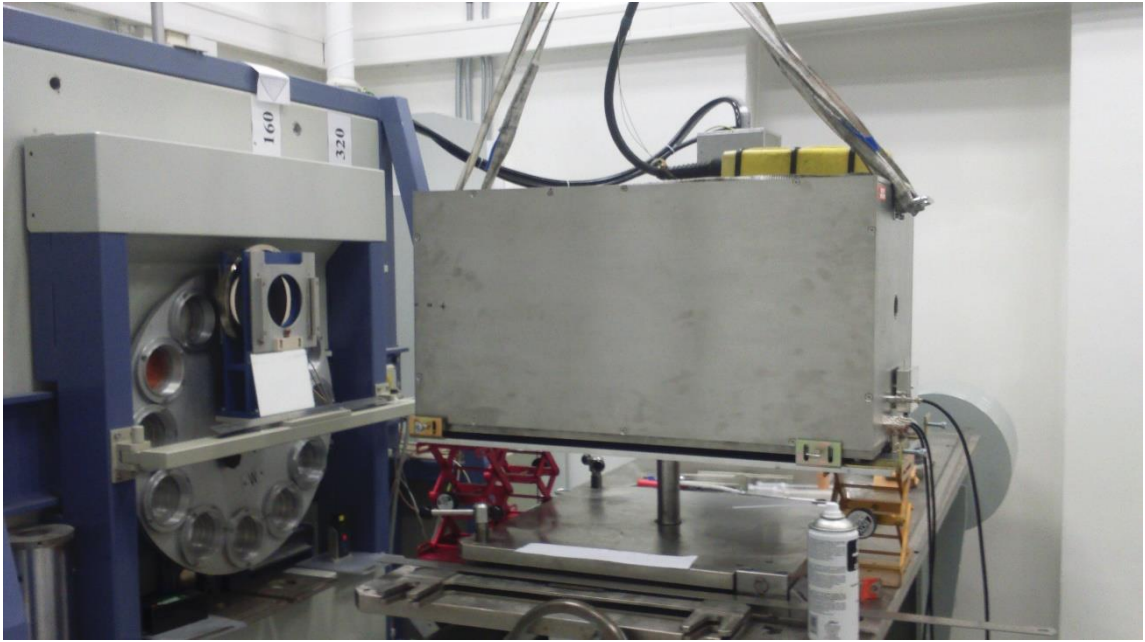

**Figure 1:** Free-air ionization chamber placed in front of the X-ray generator on the positioning table and ready for the measurement. Connected cables and the voltage divider box are visible on the rear panel.

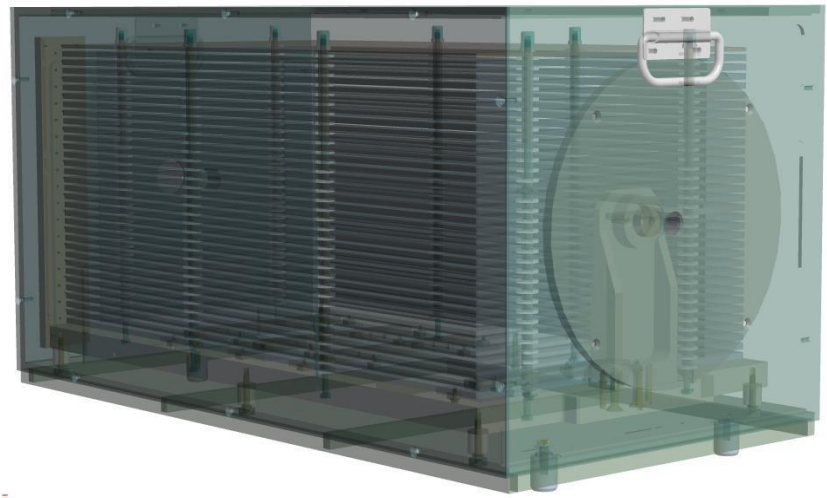

**Figure 2:** Position of the inner chamber in the outer case (CAD visualisation).

The FAC was designed with four collecting electrodes which allow different measurement modes (e.g. measurement of attenuation in air for low-energy X-rays, cross-checking of measurements). The first collecting electrode is in a distance of 100 mm from the aperture which is the same distance as in CMI free-air chamber for low-energy X-rays. It is expected to use both FACs for low energy X-rays in order to cross-check the measurement results. Due to the irradiation room layout it will not be possible to use the large FAC for mammography qualities, but it will be definitely possible for narrow spectra N10 – N30 as they are realized on the same X-ray machine as the N40 to N300 qualities. The second collecting electrode is in a distance of 320 mm where the dose from secondary electrons is fully delivered for photon energies up to 150 keV (Figure 3). The third collecting electrode is in a distance of 420 mm, so the distance between the second and the third electrode is again 100 mm. Ratio of ionization currents measured using these two electrodes should allow to experimentally determine the attenuation in air for the low-energy FAC. Fourth collecting electrode is in a distance of 540 mm in order to maximize the dose delivered by electrons. In this distance the charge particle equilibrium is virtually achieved for electrons with the initial energy of 250 keV only, however for 300 keV electrons the charge particle equilibrium is as high as 99.9%. Distance of the fourth electrode should be preferably larger, but this is a compromise required by our target values of outer dimensions and mass of the FAC. However, contribution of electrons between 250 keV to 300 keV to the air kerma is approximately 50% in the N300 spectrum, so the expected correction for non-delivered energy is less than 0.05%. Simultaneous measurement on pairs of collecting electrodes (selected depending on the energy of photons) should allow a cross-checking of measurement results. Disadvantage of this design with four collectors is smaller measuring volume due to shorter length of electrodes.

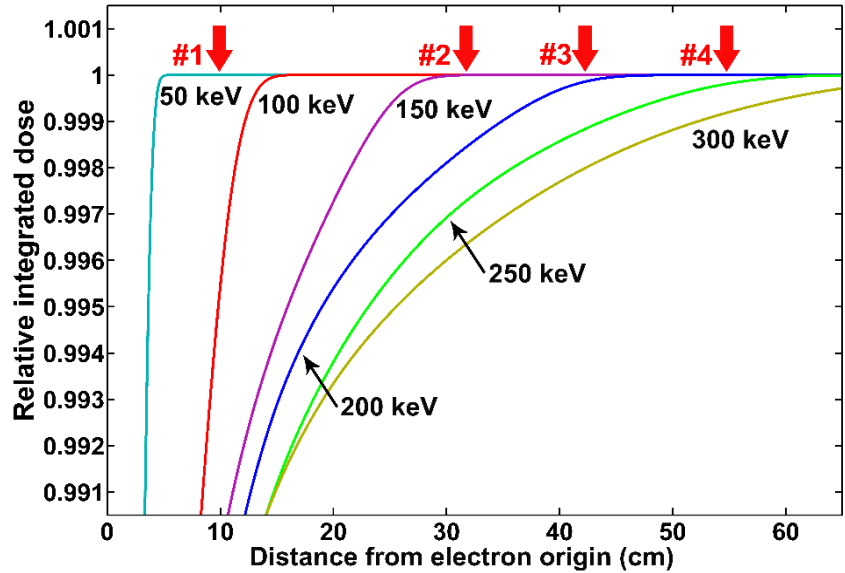

**Figure 3:** Calculated relative integrated absorbed dose from secondary electrons as a function of their primary energy and the distance from their origin. Red arrows indicate the positions of four collecting electrodes.

Basic part of the FAC design is the base guarding electrode in which four collecting electrodes are very precisely inserted using supporting insulator pads in a such way, that the gaps between the guarding electrode and individual collecting electrodes are very well defined (Figures 4 and 5). The guarding and collecting electrodes are made of aluminium alloy; supporting pads are made of isolating material based on glass filaments. A holder (made of aluminium alloy) is fitted on the guarding electrode for cylindrical tungsten aperture with entrance opening defining the cross section of X-ray beam entering the chamber. The base guarding electrode, collecting electrodes, supporting pads and aperture holder create a basic assembly, which is not supposed to be dismantled at usual circumstances, because the dimensions and mutual position of these parts define the basic parameters of FAC.

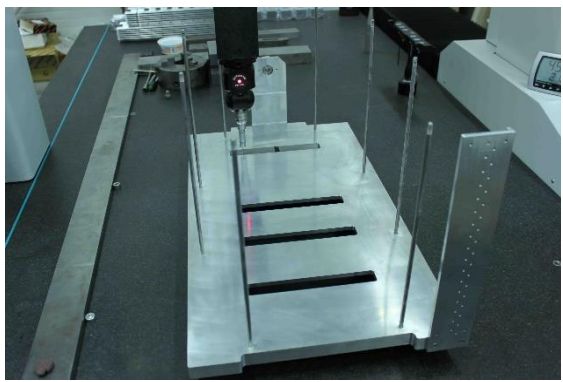
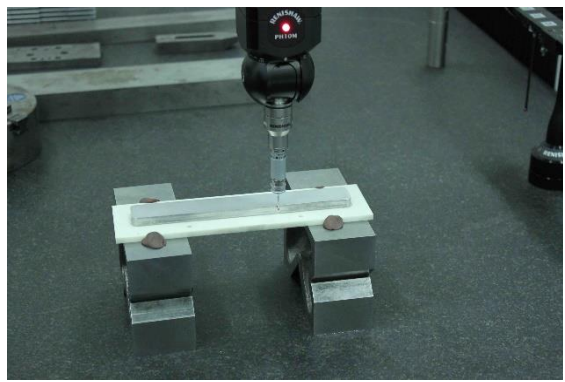

**Figure 4:** Base electrode with the aperture holder and openings for collecting electrodes, placed in a coordinate measurement machine (CMM).

**Figure 5:** Measurement of collecting electrode dimensions using CMM.

The next part is a box of guarding rings. The box consists of a set of 33 rectangular rings made of aluminium alloy sheets with a thickness of 6 mm (Figures 6 and 7). The guarding rings are separated by 3 mm thick glass pads. Guarding rings and glass pads are placed on eight electrically isolated steel rods fixed in the base guarding electrode. The guarding rings are cut to create entrance and output openings for X-ray beam (Figures 8 and 9). The box of guarding rings is closed by a high-voltage electrode made of aluminium alloy.

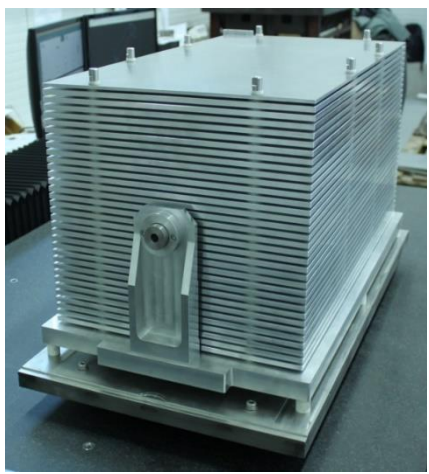
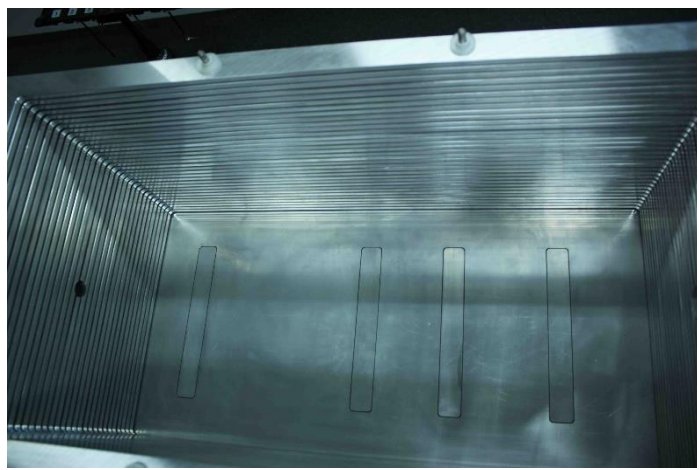

**Figure 6:** Completed free-air chamber (without outer case). In front: cylindrical tungsten aperture in its holder.

**Figure 7:** View into the inner chamber to collecting electrodes (HV electrode removed).

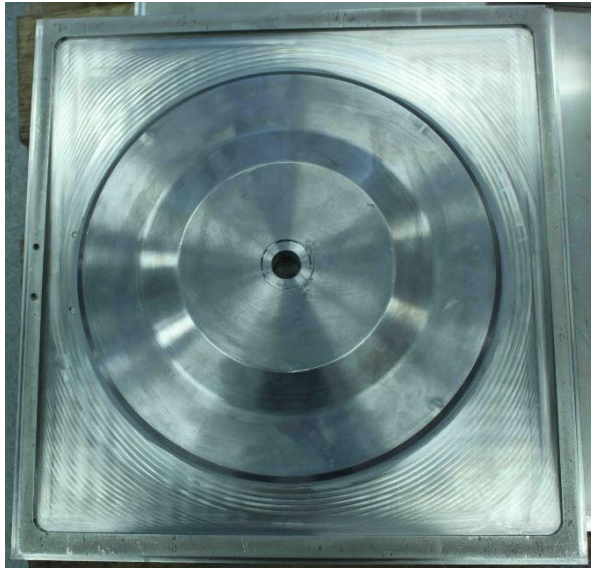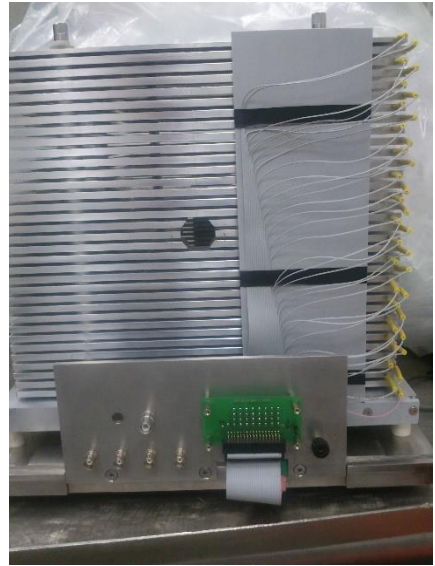

**Figure 8:** Inner part of the front wall with the lead radiation shielding.

**Figure 9:** Back wall of the chamber with the voltage divider and connectors for high voltage and measurement of ionizing current (outer case removed).

Both parts, the guarding electrode with collecting electrodes and the box of guarding rings, create an assembly, where the area $A$ of the entrance opening and the distance $l$ of the collecting electrode from the aperture are defined and where a homogeneous electric field can be created. This assembly is placed in outer case using distance poles. The stainless steel outer case of FAC ensures an electromagnetic shielding, mechanical protection of the chamber and allows handling with a chamber, but it is not used as a radiation shielding of the FAC. Monte Carlo simulations showed that for photon beam diameters of 60 mm to 100 mm only the radiation shielding on the front panel is necessary. Therefore a circular lead shielding with the diameter of 300 mm and a variable thickness of 25 mm and 15 mm was added to the front panel.

The collecting electrodes are connected to the connector on the back wall of the chamber. The guard rings and the high voltage electrode are connected to the voltage divider ensuring an equidistant voltage division. The divider consists of a series of 10 MΩ resistors and it is placed outside the chamber in order to avoid influence to the inner temperature of the FAC. High voltage is connected using a MHV connector on the back face of the outer case (Figures 1 and 9).

Basic dimensions of the FAC are summarized in Table 1.

**Table 1:** Basic parameters of the FAC.

| Parameter | Value | | | |
|---|---|---|---|---|
| Inner cavity dimensions (mm$^3$) | 300×300×600 | | | |
| Aperture diameter (mm) | 10.169 | | | |
| | **Electrode No.** | | | |
| | 1 | 2 | 3 | 4 |
| Distances from the aperture to the centre of collecting electrode (mm) | 100 | 320 | 420 | 540 |
| Collecting electrode lengths (mm) | 24.988 | 24.954 | 24.987 | 24.986 |
| Air gaps (total width of both sides; mm) | 1.025 | 1.052 | 1.019 | 1.027 |
| Measuring volumes (cm$^3$) | 2.0711 | 2.0694 | 2.0707 | 2.0710 |
| Calibration coefficient, for 101.325 kPa and 20°C (Gy/C) | 1.36117.10$^7$ | 1.36227.10$^7$ | 1.36139.10$^7$ | 1,36123.10$^7$ |

## 2.3 FAC operation

The FAC was designed and manufactured in 2013 and 2014. It was put into operation in 2015 and its operational properties were checked, however some issues were identified especially with the insulators of the fixing steel rods or with the FAC alignment against the tube focus. In 2016 the insulators were replaced, so the working high voltage could be set to 4500 V and new values of $k_{pol}$ and $k_{sat}$ were determined for this high voltage.

The FAC positioning table was improved and a procedure was developed to align the FAC with the photon beam correctly by a step-by-step positioning of the rear panel of FAC against the X-ray beam visualised by an intensifying foil (Figure 10).

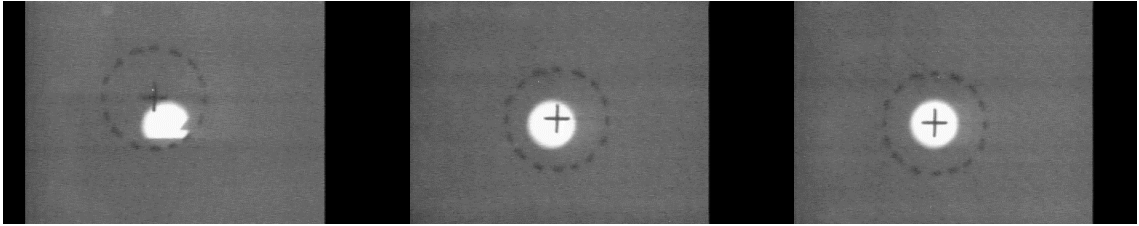

**Figure 10:** Iterative procedure of FAC alignment: The dashed circle represents the rear opening of FAC, the bright circle is the visualised front opening. Goal is to align these two circles.

## 2.4 Monte Carlo simulations

The development of the FAC was supported by Monte Carlo (MC) simulations performed using the general-purpose MC code MCNPX$^{TM}$ in version 2.7.E [8]. The complex MC model reflecting the final version of the FAC (Figure 11) was used for determination of correction factors $k_{att}$, $k_{tr}$, $k_{sc}$, $k_{fl}$, $k_{loss}$, $k_{br}$, $k_{dtr}$, and $k_{dsc}$. The procedure for determination of the correction factors is presented in [1] and [9].

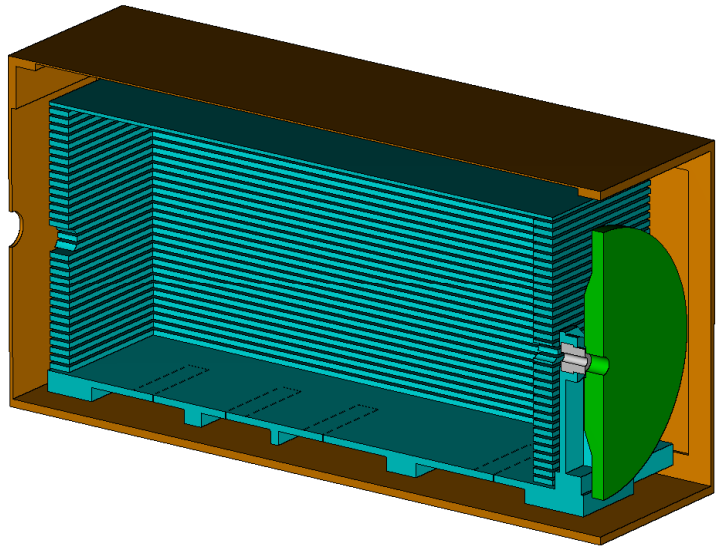

**Figure 11:** Visualization of the FAC Monte Carlo model.

Figure 12 presents an example of visualization of electron tracks inside the FAC, for N150 beam quality (150 kV). Tracks of primary electrons originating in the photon beam clearly visualize beam diameter inside the FAC.

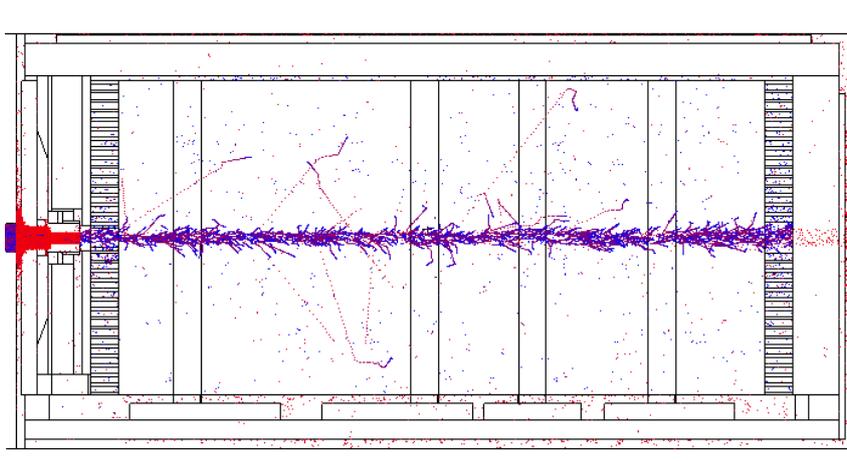

**Figure 12:** Simulation of electron tracks in the chamber for N150 quality using the MCNPX code.

## 3. Results

### 3.1 Total correction factors

The final values of the total correction factors were determined for all four collecting electrodes, for spectral fluence distributions of source photons taken from [10]. Table 2 summarizes the total corrections for all four collecting electrodes. The presented total corrections include only the individual corrections obtained from Monte Carlo simulations, i.e., it is a product of $k_{att}$, $k_{sc}$, $k_{fl}$, $k_{loss}$, $k_{br}$, $k_{dtr}$, and $k_{dsc}$.

**Table 2:** FAC total correction factor from corrections obtained from MC simulations, for 5 cm beam radius.

| Quality | FAC total MC correction factor (beam radius: 5 cm) | | | |
|---|---|---|---|---|
| | collector #1 | collector #2 | collector #3 | collector #4 |
| **N40** | 0.997 | 1.005 | 1.009 | 1.014 |
| **N60** | 0.996 | 1.001 | 1.004 | 1.007 |
| **N80** | 0.995 | 0.999 | 1.001 | 1.004 |
| **N100** | 0.996 | 0.999 | 1.001 | 1.004 |
| **N120** | 1.000 | 1.004 | 1.005 | 1.008 |
| **N150** | 1.004 | 1.010 | 1.012 | 1.014 |
| **N200** | 1.000 | 1.008 | 1.010 | 1.011 |
| **N250** | 0.996 | 1.003 | 1.004 | 1.006 |
| **N300** | 0.993 | 0.999 | 1.000 | 1.003 |

### 3.2 Uncertainty budget

The uncertainty budget of the FAC measured value is presented in Table 3.

**Table 3:** The FAC measured value uncertainty budget.

| Relative standard uncertainty [%] | $u_A$ | $u_B$ |
|---|---|---|
| Ionisation current | 0.15 | 0.23 |
| Positioning | - | 0.07 |
| Volume | - | 0.06 |
| Correction factors | - | 0.35 |
| Physical constants | - | 0.35 |
| Total | 0.15 | 0.55 |
| | | 0.57 |
| $U_c$ (k=2) | | 1.1 |

### 3.3 Verification of $k_{att}$

A preliminary rough check of the MC-calculated attenuation correction factors, $k_{att}$, of the low-energy FAC was performed by comparison to the ratio of corrected ionization currents $I_2$ and $I_3$ measured simultaneously by 2$^{nd}$ and 3$^{rd}$ collecting electrode of the medium-energy FAC. The results are presented in Table 4.

**Table 4:** Comparison of measured and simulated $k_{att}$, at 10 cm air thickness.

| Quality | Tube current [mA] | $k_{att}$ Monte Carlo | $I_2/I_3$ experiment |
|---|---|---|---|
| **N10** | 30 | 1.081 | 1.090 |
| **N15** | 30 | 1.030 | 1.032 |
| **N20** | 30 | 1.016 | 1.014 |
| **N25** | 30 | 1.009 | 1.007 |
| **N30** | 40 | 1.005 | 1.004 |
| **N40** | 20 | 1.006 | 1.007 |

The values seem to be in a quite good agreement, however this procedure still needs more detailed analysis and improvement.

### 3.4 Comparison to secondary standard

An indicative functional test of the FAC was performed by comparison to the secondary ionisation chamber Exradin A4 (Standard Imaging, USA) traceable to the primary standard of BEV, Austria. The results are stated in Table 5.

**Table 5:** Comparison of FAC to the secondary chamber Exradin A4.

| Quality | Tube current [mA] | A4 [Gy/s] | FAC [Gy/s] | Deviation [%] |
|---|---|---|---|---|
| **N40** | 20 | 2.420E-05 | 2.403E-05 | -0.7% |
| **N60** | 20 | 3.750E-05 | 3.752E-05 | 0.1% |
| **N80** | 20 | 2.054E-05 | 2.051E-05 | -0.2% |
| **N100** | 30 | 1.450E-05 | 1.445E-05 | -0.3% |
| **N120** | 25 | 1.331E-05 | 1.345E-05 | 1.0% |
| **N150** | 10 | 4.074E-05 | 4.080E-05 | 0.1% |
| **N200** | 20 | 3.018E-05 | 3.068E-05 | 1.7% |
| **N250** | 15 | 2.318E-05 | 2.312E-05 | -0.3% |
| **N300** | 13 | 1.915E-05 | 1.940E-05 | 1.3% |

Uncertainty of the air kerma rate value measured by the secondary chamber is 1.4% (k=2).

## 4. Conclusions

A free air chamber for the standardization of air kerma quantity in medium-energy X-ray beams up to 300 keV was built in the Czech Metrology Institute. Necessary correction factors were determined and whole measuring chain including the application of correction factors was tested by a comparison to the calibrated secondary ionizing chamber. The chamber is considered ready for a key comparison.

## Acknowledgments

This work was supported by the European Metrology Programme for Innovation and Research (EMPIR) joint research project 14RPT04 "Absorbed dose in water and air" (Absorb; http://absorb-empir.eu/) which has received funding from the European Union. The EMPIR initiative is co-funded by the European Union's Horizon 2020 research and innovation programme and the EMPIR Participating States.